\documentclass[aps,pra,final,10pt,twocolumn,superscriptaddress,nofootinbib,flushbottom]{revtex4-1}%
\usepackage{amssymb,amsmath,amsfonts}
\usepackage{graphicx,color}
\usepackage[squaren]{SIunits}
\usepackage[english]{babel}

\newcommand{\mathbfup}[1]{\mathrm{\mathbf{#1}}}%
\newcommand{\AU}[1]{{\fignumfont #1}}%
\newcommand{\dif}{\mathrm{d}}%

\makeatletter
\def\frutiger{cmss10 }
\def\frutigerbold{cmssbx10 }
=\frutigerbold at 14pt
=\frutiger at 8pt
=\frutigerbold at 12pt
=\frutigerbold at10pt
=\frutigerbold at 8pt
=\frutiger at 8pt
=\frutigerbold at 31pt
\def\@caption@tabnum@sep{\figtextfont{{ }{\bf\textbar}{ }}}%
\def\fnum@table{{\bf\tablename~\thetable}}

\def\@caption@fignum@sep{\figtextfont{{ }{\bf\textbar}{ }}}%
\def\fnum@figure{{\fignumfont\figurename~\thefigure}}
\renewenvironment{figure}{\@float{figure}\def\textbf##1{{\fignumfont ##1}}\def\bf{\fignumfont}}{\end@float}
\renewenvironment{figure*}{\@dblfloat{figure}\def\textbf##1{{\fignumfont ##1}}\def\bf{\fignumfont}}{\end@dblfloat}
\def\@startsection#1#2#3#4#5#6{%
\if@noskipsec\leavevmode\fi
\par\@tempskipa #4\relax
\@afterindenttrue
\ifdim\@tempskipa <\z@
\@tempskipa -\@tempskipa \@afterindentfalse
\fi\if@nobreak\everypar{}%
\else\addpenalty\@secpenalty\addvspace\@tempskipa\fi
\@ifstar{\@ssect{#3}{#4}{#5}{#6}}{\@dblarg{\@sect{#1}{#2}{#3}{#4}{#5}{#6}}}}
\def\@sect#1#2#3#4#5#6[#7]#8{%
\ifnum #2>0
\let\@svsec\@empty
\else\refstepcounter{#1}\protected@edef\@svsec{\@seccntformat{#1}\relax}\fi
\@tempskipa #5\relax
\ifdim\@tempskipa>\z@
\begingroup#6{\@hangfrom{\hskip #3\relax\@svsec}%
\interlinepenalty \@M #8\@@par}\endgroup
\csname #1mark\endcsname{#7}%
\addcontentsline{toc}{#1}{%
\ifnum #2>\c@secnumdepth\else
\protect\numberline{\csname the#1\endcsname}\fi #7}%
\else\def\@svsechd{#6{\hskip #3\relax
\@svsec #8\ifnum#2=2.\fi}%
\csname #1mark\endcsname{#7}%
\addcontentsline{toc}{#1}{%
\ifnum #2>\c@secnumdepth \else
\protect\numberline{\csname the#1\endcsname}\fi #7}}%
\fi\@xsect{#5}}
\renewcommand\section{\@startsection {section}{1}{\z@}%
{-10pt \@plus -1ex \@minus -.2ex}{.5ex }{\fontsize{12}{10}\bfseries\sectionfont}}
\renewcommand\subsection{\@startsection{subsection}{2}{\z@}%
{10pt\@plus 1ex \@minus .2ex}{-1ex \@plus .2ex}{\normalfont\large\bfseries\subsectionfont}}
\def\frontmatter@title@format{\titlefont\centering}%
\def\frontmatter@title@below{\addvspace{-5pt}}%
\def\dropcap#1{\setbox1=\hbox{\dropcapfont\uppercase{#1}\hskip1pt}
\hangindent=\wd1 \hangafter-2 \noindent\llap{\vbox to0pt{\vskip-7pt\copy1\vss}}}

\newcommand*\bib@heading{%
  \section{\refname}%
  \fontsize{8}{10}\selectfont
}
\newcommand*\@openbib@code{%
      \advance\leftmargin\bibindent
      \itemindent -\bibindent
      \listparindent \itemindent
      \parsep \z@
}%
\newdimen\bibindent
\renewcommand{\footnoterule}{%
  \kern -5pt 
  \hrule width 1in height 0.5pt
  \kern 5.5pt
}
\makeatother

\begin{document}%
\title{Scalar {\Large$\boldsymbol{\phi^{4}}$} field theory for active-particle phase separation}%

\author{Raphael Wittkowski}
\email[Corresponding author: ]{raphael.wittkowski@ed.ac.uk}
\affiliation{SUPA, School of Physics and Astronomy, University of Edinburgh, Edinburgh EH9 3JZ, United Kingdom}

\author{Adriano Tiribocchi}
\affiliation{SUPA, School of Physics and Astronomy, University of Edinburgh, Edinburgh EH9 3JZ, United Kingdom}

\author{Joakim Stenhammar}
\affiliation{SUPA, School of Physics and Astronomy, University of Edinburgh, Edinburgh EH9 3JZ, United Kingdom}

\author{Rosalind J. Allen}
\affiliation{SUPA, School of Physics and Astronomy, University of Edinburgh, Edinburgh EH9 3JZ, United Kingdom}

\author{Davide Marenduzzo} 
\affiliation{SUPA, School of Physics and Astronomy, University of Edinburgh, Edinburgh EH9 3JZ, United Kingdom}

\author{Michael E. Cates} 
\affiliation{SUPA, School of Physics and Astronomy, University of Edinburgh, Edinburgh EH9 3JZ, United Kingdom}

\date{\today}

\begin{abstract}%
Recent theories predict phase separation among orientationally disordered active particles whose propulsion speed decreases rapidly enough with density. 
Coarse-grained models of this process show time-reversal symmetry (detailed balance) to be restored for uniform states, but broken by gradient terms; hence detailed-balance violation is strongly coupled to interfacial phenomena.
To explore the subtle generic physics resulting from such coupling we here introduce `Active Model B'. This is a scalar $\phi^{4}$ field theory (or phase-field model) that minimally violates detailed balance via a leading-order square-gradient term. We find that this additional term has modest effects on coarsening dynamics, but alters the static phase diagram by creating a jump in (thermodynamic) pressure across flat interfaces. Both results are surprising, since interfacial phenomena are always strongly implicated in coarsening dynamics but are, in detailed-balance systems, irrelevant for phase equilibria.  
\end{abstract}

\maketitle
\newpage

\dropcap{M}uch recent research has addressed active materials, whose constituent particles violate microscopic time-reversal symmetry (TRS) by continuously converting fuel into motion. Systems with medium- or long-range orientational order, such as swarms of rodlike bacteria \cite{WensinkDHDGLY2012}, can then be viewed as active liquid crystals, which are successfully described by adding minimal active terms to the established continuum equations for liquid crystal hydrodynamics. The resulting field theories involve either vector or tensor order parameters describing the local state of orientational order \cite{Ramaswamy2010,MarchettiJRLPRS2013}.

However, an important alternative paradigm addresses \textit{isotropic} bulk phases of active (i.e., self-propelled) colloidal particles. These can be natural microorganisms such as bacteria or algae \cite{Cates2012,MarchettiJRLPRS2013}, or synthetic  microswimmers \cite{HowseJRGVG2007,EbbensH2010,ThutupalliSH2011,VolpeBVKB2011,PalacciSSPC2013}. 
In many such cases, the dynamics is approximately described by having a fixed relaxation time $\tau$ for the direction of self-propelled motion, but a non-trivial dependence of the propulsion speed $v(\rho)$ on the particle density $\rho$. 
If $\dif\ln(v)/\dif\ln(\rho)<-1$, steady-state phase separation can be shown to arise through a positive feedback mechanism in which a slowing of the particles leads to their accumulation, and vice versa \cite{TailleurC2008,CatesT2013}. 
In biological systems, such slowing could arise through various causes such as a coupling of the quorum-sensing response of bacteria to their motility \cite{LiuEtAl2011,Cates2012}. For synthetic swimmers, which are often modelled as `active Brownian particles' (ABPs) whose swimming direction rotates by diffusion, an effective decrease in $v$ at high density can result instead from collisional interactions \cite{FilyM2012,StenhammarTAMC2013}.

This scenario of active-particle phase separation is now fully established in simulation studies \cite{CatesMPT2010,ThompsonTCB2011,FilyM2012,RednerHB2013,StenhammarTAMC2013,WysockiWG2014}, and partly confirmed by experiments  \cite{ButtinoniBKLBS2013,LiuDRdJHRvdK2013}. 
It cannot happen in systems with microscopic TRS for which steady-state accumulation by slowing is forbidden. 
Indeed, TRS of the steady state, also known as `the principle of detailed balance', ensures that the equilibrium density is controlled solely by conservative interparticle forces (via the Boltzmann distribution) and not by any density-dependent kinetic coefficients \cite{TailleurC2008,Schnitzer1993}. 
For example, in an isothermal suspension of Brownian colloidal particles, the diffusivity decreases strongly at high density, with no effect on the phase diagram. In such `passive' colloids, phase separation requires attractive interactions. 

For passive colloidal systems, the resulting phase-separation kinetics is well described by dynamical field theories involving a conserved scalar order-parameter field $\phi$, linearly related to the local density of colloids. Simplifying the underlying free energy to a quartic polynomial in $\phi$ with square-gradient terms, and assuming local diffusive dynamics, then gives a theory called `Model B'. Such scalar $\phi^{4}$ field theories (or phase-field models) have played a pivotal role in understanding phase separation in systems with TRS \cite{HohenbergH1977,ChaikinL1995}. This applies particularly when noise terms are neglected, creating a mean-field model that accurately captures the long-time dynamics of phase separation, which is dominated by the deterministic motion of sharply defined interfaces. In that limit, Model B becomes the simplest form of Cahn-Hilliard equation \cite{CahnH1958}, capturing a celebrated universal result, $L\sim t^{1/3}$, for the dependence of the domain size $L$ on time $t$ \cite{Bray1994}. 

Despite the deep distinction between active and passive phase separation, explicit coarse-graining of the active dynamics at large scales establishes a partial mapping between the two cases. This mapping was found first for models of swimming bacteria with discrete reorientations \cite{TailleurC2008}, but later extended to ABPs \cite{CatesT2013}.
At zeroth order in spatial gradients, which is equivalent to considering only systems of uniform density, the mapping allows a bulk free energy to be constructed whose instabilities are those of the active system \cite{TailleurC2008,CatesT2013}. 
(In what follows, all thermodynamic quantities, such as pressure and bulk chemical potential, refer to those calculated within this mapping.) 
Detailed balance, while absent microscopically, re-emerges to this order. Recently, however, we studied the leading-order gradient terms and found these to break detailed balance once again \cite{StenhammarTAMC2013}. This creates a new class of models in which the breaking of TRS is intimately linked to the physics of interfaces. 

This feature distinguishes such models from others that address Cahn-Hilliard-like diffusive instabilities in systems without detailed balance (see, e.g., refs.\ \cite{KhainS2008,Murray2008,Risken1996,WatsonN2006,WatsonORD2003}). The latter encompass many physical processes but are often too complex for their fundamental physics to be understood. In elucidating the generic physics of active-particle phase separation, it is therefore important to focus on the simplest model of the required structure.

In this article we present and analyse just such a model. To create our new model we add the simplest `nonintegrable' (as defined below) gradient term to what is otherwise the standard field theory for locally diffusive phase separation, namely Model B \cite{HohenbergH1977,ChaikinL1995}. 
The chosen gradient term breaks detailed balance in the standard, passive Model B, which implies that the resulting \textit{Active Model B} cannot be derived from any free-energy functional.  

Below we report simulations of phase kinetics with Active Model B which echo results found previously using a more elaborate continuum model inspired by ABP simulations \cite{StenhammarTAMC2013}. We find that the nonintegrable term does not greatly alter the dynamical fate of the system. This is remarkable, since coarsening dynamics is controlled by interfacial tension, which vanishes without the gradient terms and so ought to be sensitive to their form. It also presents a paradox when confronted by another new result: 
Active Model B admits no static domain-wall solution connecting two bulk regions whose chemical potential $\mu_{0}\equiv\dif f_{0}/\dif\phi$ with the bulk free-energy density $f_{0}$ takes the value set by the common tangent construction. (As recalled below, this construction holds at coexistence, independent of gradient terms, for all systems with detailed balance.) We explain the resulting `uncommon tangent' result in terms of an activity-induced analog of Laplace pressure that arises even across flat interfaces. This insight allows us to explain why active coarsening dynamics, en route to full phase separation, remain similar to that of the traditional Passive Model B.

\section{Results}%
\subsection{Active Model B} 
In line with the principles outlined above, we adopt the following dynamics for a conserved scalar order-parameter field $\phi(\mathbfup{r},t)$ at position $\mathbfup{r}$ and time $t$ in $d$ dimensions:
{\allowdisplaybreaks
\begin{align}%
\dot{\phi} &= -\nabla\!\cdot\!\mathbfup{J} \;, \label{phidot}\\
\mathbfup{J} &= -\nabla \mu +\mathbfup{\Lambda} \;, \label{J}\\
\mu &= -\phi +\phi^{3} -\nabla^{2}\phi + \lambda(\nabla\phi)^{2} \;. \label{mu}
\end{align}}%
All quantities are made dimensionless by using `natural units'; these are $v(0)\tau$ for length and the orientational relaxation time $\tau$ for time, where $v(0)$ is the swim speed of an isolated particle. 
The composition variable $\phi$ is related to the number density $\rho(\mathbfup{r},t)$ of active particles by a linear transform $\phi=(2\rho-\rho_{\mathrm{H}}-\rho_{\mathrm{L}})/(\rho_{\mathrm{H}}-\rho_{\mathrm{L}})$, where $\rho_{\mathrm{H}}$ and $\rho_{\mathrm{L}}$ are the densities of high- and low-density coexisting phases, respectively, as calculated for example from $v(\rho)$ by the methods of refs.\ \cite{TailleurC2008,CatesT2013}.

Here eq.\ \eqref{phidot} expresses conservation of $\phi$, while eq.\ \eqref{J} states that its mean current $\mathbfup{J}-\mathbfup{\Lambda}$ is proportional to the gradient of a nonequilibrium chemical potential $\mu$ obeying eq.\ \eqref{mu} with a constant $\lambda$. 
Our nomenclature for $\mu$ is self-explanatory: even beyond equilibrium, the chemical potential is the quantity whose gradient causes the mean current. 
The vector $\mathbfup{\Lambda}$ is a Gaussian white noise whose variance we take to be constant. This follows standard practice in Passive Model B although in reality the variance is density-dependent, as calculated explicitly for active particles in ref.\ \cite{TailleurC2008}. The noise is often neglected altogether for phase-separation studies \cite{Bray2001} and we generally ignore it below. 

In eq.\ \eqref{mu}, $\mu=\mu_{0}+\mu_{1}$ is the sum of bulk and gradient contributions. The bulk part is chosen as the usual Passive Model B form, $\mu_{0}=-\phi+\phi^{3}$, so that at zeroth order in a gradient expansion, our Active Model B shares with its passive counterpart the bulk free-energy density of a symmetric $\phi^{4}$ field theory, $f_{0}=-\phi^{2}/2+\phi^{4}/4$. Note that the phase separation, driven by the negative linear term in $\mu_{0}$, can arise from activity alone with no need for attractive interactions.
The gradient term $\mu_{1}=\mu^{\mathrm{P}}_{1}+\mu^{\mathrm{A}}_{1}$ is the sum of two further terms. The first is an integrable or `passive' piece $\mu^{\mathrm{P}}_{1}$, which can be written as a functional derivative of some free energy $\int\!f_{1}\,\dif^{d}r$, while the second is an active part $\mu^{\mathrm{A}}_{1}$, which cannot.  

We now make the standard Passive Model B choice, $f_{1}=(\nabla\phi)^{2}/2$, so that $\mu^{\mathrm{P}}_{1}=-\nabla^{2}\phi$. 
In the passive case, choosing the total free-energy density $f=f_{0}+f_{1}$, in which $f_{0}$ is supplemented by the simplest square-gradient term $f_{1}=(\nabla\phi)^{2}/2$, captures all universal aspects of the underlying physics, while allowing vastly simpler analysis of interfacial structure and dynamics than would a more realistic choice of $f$. The same advantages hold for our Active Model B. 
For the nonintegrable term we write $\mu^{\mathrm{A}}_{1}=\lambda(\nabla\phi)^{2}$; the constant $\lambda$ is a parameter of the model. This is the simplest addition to $\mu$, at second order in gradients, that \textit{cannot} be derived from a free-energy or Lyapunov functional. 
Note that this property is the definition of `nonintegrable' for the purposes of the current paper. 
 
Explicit coarse-graining of the dynamics of ABPs \cite{StenhammarTAMC2013,BialkeLS2013} points to a specific structure of the gradient terms in eq.\ \eqref{mu} and leads to a gradient term $\mu_{1}=-\kappa(\phi)\nabla^{2}\phi$ with $\kappa(\phi) = 1+2\lambda\phi$ that combines exactly our $\lambda$ term $\mu^{\mathrm{A}}_{1}$ with an integrable part $\mu^{\mathrm{P}}_{1}$ that corresponds to $f_{1} = \kappa(\phi)(\nabla\phi)^{2}/2$.
In the units of Active Model B, the parameter $\lambda$, whose sign can be absorbed into that of $\phi$ if preferred, is then negative and of order unity for ABPs (see Supplementary Note 1 where an explicit expression for $\lambda$ is given). 
The same is true for run-and-tumble bacteria whose dynamics are almost equivalent \cite{CatesT2013}. In both cases, $\lambda$ is set primarily by the \textit{shape} of the function $v(\rho)$, parameterizing the decay with density of the mean swimming speed. Meanwhile, dimensioned parameters such as $v(0)$ and $\tau$ serve to set conversion factors between Active Model B units and laboratory ones.

Replacement of the derived coarse-grained form of $f_1$ with one having constant $\kappa = 1$ is standard practice for Passive Model B, just as one replaces a complicated coarse-grained $f_0$ with the standard form $f_0=-\phi^2/2+\phi^4/4$. 
(This is how Passive Model B comes to describe a wide range of microscopic models.)
In Active Model B, we make exactly the same simplifications for all the integrable terms, while capturing new physics with the minimal, leading-order TRS-breaking term $\mu^{\mathrm{A}}_{1}=\lambda(\nabla\phi)^{2}$. Furthermore, we show in Supplementary Note 1 that all possible leading-order current contributions, up to third order in $\nabla$ and second order in $\phi$, are given by combining some choice of free-energy density $f$ with this choice of $\mu^{\mathrm{A}}_{1}$.

Active Model B's nonintegrable term somewhat resembles one arising in the celebrated Kardar-Parisi-Zhang equation for nonlinear interfacial diffusion \cite{KardarPZ1986}, which was constructed on similar minimalist grounds, and likewise supported by direct contact with microscopic arguments for specific examples.  

\begin{figure}
\centering
\includegraphics[width=\columnwidth]{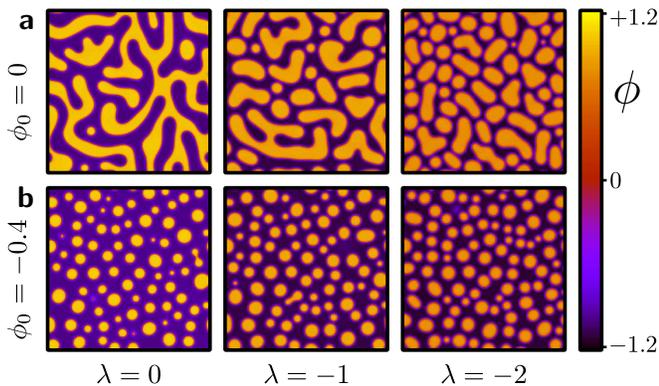}
\caption{\label{Fig.1}\AU{Transient domain structures.} Snapshots of evolving phase separation in two dimensions at time $t=2000$ and $\lambda=0,-1,-2$ for (\textbf{a}) symmetric ($\phi_{0}=0$) and (\textbf{b}) asymmetric ($\phi_{0}=-0.4$) quenches. The plots shown have dimensions $256\times 256$.}
\end{figure}
\begin{figure*}
\centering
\includegraphics[width=\linewidth]{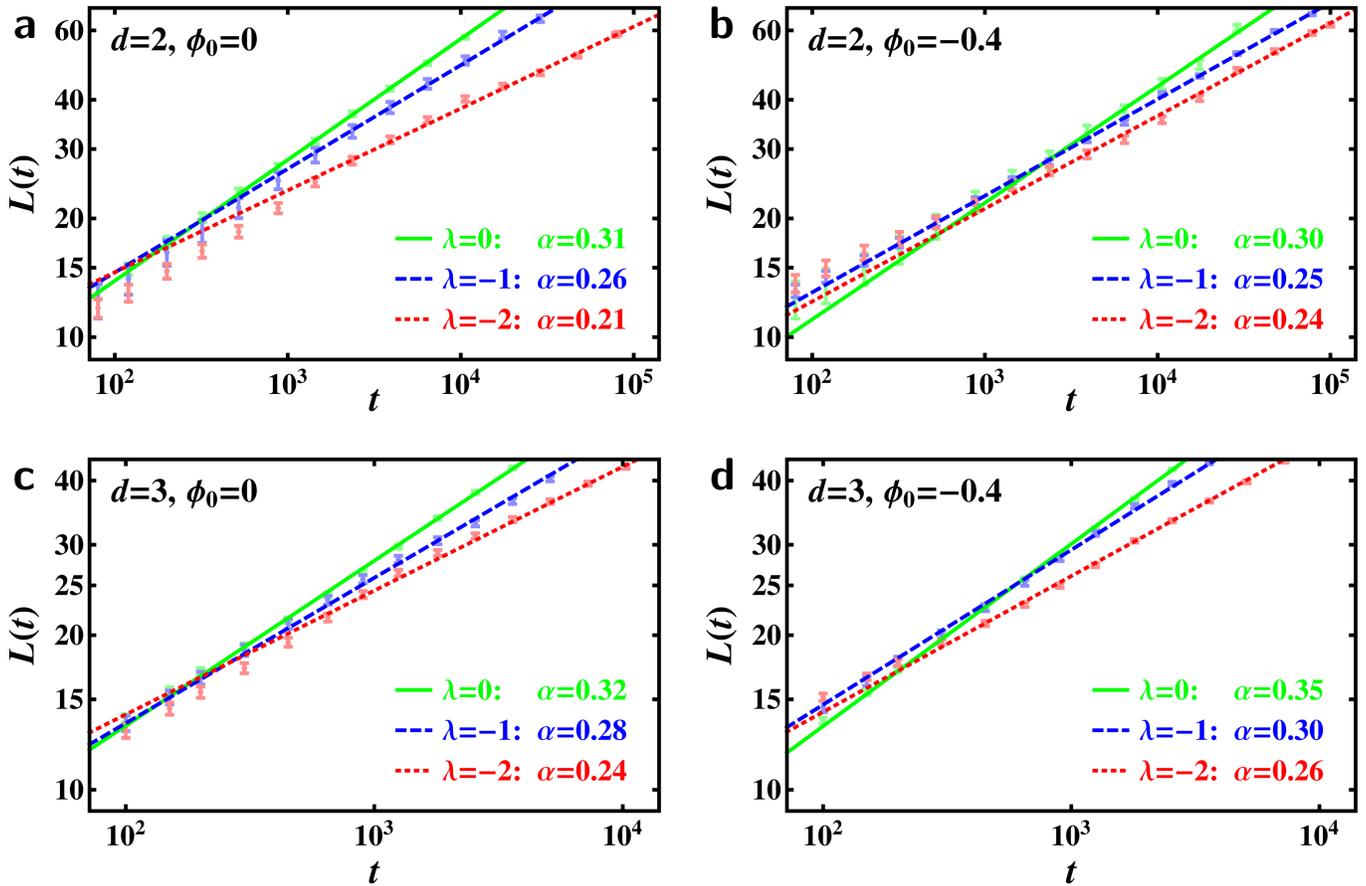}
\caption{\label{Fig.2}\AU{Time evolution of the characteristic domain size.} Numerical results (points with error bars) and least-squares fits (straight lines) for the domain length scale $L(t)$ for different $\lambda$ and average density $\phi_{0}$ in both  two and three dimensions: (\textbf{a}) $d=2$, $\phi_{0}=0$; (\textbf{b}) $d=2$, $\phi_{0}=-0.4$; (\textbf{c}) $d=3$, $\phi_{0}=0$; (\textbf{d}) $d=3$, $\phi_{0}=-0.4$. Notice the apparent downward drift with increasing $\lvert\lambda\rvert$ of the exponent $\alpha$ from the well-known value $\alpha=1/3$ of Passive Model B. The error bars denote the standard
deviations of the numerical data from the corresponding fit curves. The definition of $L(t)$ is given in the Methods section.}
\end{figure*}

\subsection{Dynamics of Active Model B} 
We have explored the dynamics of Active Model B numerically for both symmetric ($\int\!\phi\,\dif^{d}r=0$) and asymmetric ($\int\!\phi\,\dif^{d}r\neq 0$) quenches in which a uniform initial state, with slight noise, is evolved by eqs.\ \eqref{phidot}-\eqref{mu}. Our findings are in keeping with a previous study which showed that the nonintegrability leads to at most quantitative but not qualitative changes in coarsening dynamics \cite{StenhammarTAMC2013}. This applies here also, bearing in mind that the $\lambda$ term also breaks the $\phi\to-\phi$ symmetry of a symmetric quench. (In two dimensions this results in a droplet rather than a bicontinuous domain structure, resembling a slightly asymmetric quench.) Figure \ref{Fig.1} shows snapshots of evolving domain structures for $\lambda=0,-1,-2$ with mean initial order-parameter fields corresponding to a symmetric quench (with average density $\phi_{0}\equiv\langle\phi(\mathbfup{r},0)\rangle=0$) and a dilute quench ($\phi_{0}=-0.4$). The case of a dense quench ($\phi_{0} = +0.4$) is not very different except for a slightly slower coarsening rate. As noted above, results for positive $\lambda$ can be generated by reversing the sign of $\phi$.
Corresponding data for $\lvert\lambda\rvert\le 0.1$ is not shown, but very close to the $\lambda=0$ case. 

Figure \ref{Fig.2} shows the time evolution of the characteristic domain size $L(t)\sim t^{\alpha}$ for various $\lambda$ and $\phi_{0}$ in two and three dimensions. 
Here we see some evidence for a downward drift with increasing $\lvert\lambda\rvert$ of the exponent $\alpha$ from the well-known value $\alpha=1/3$ of Passive Model B \cite{Bray2001}. To distinguish a real asymptotic shift in $\alpha$ from a $\lambda$-influenced crossover would require an exhaustive computational study \cite{KendonCPDB2001} which we leave for future work. Note however that a similar downward exponent shift was reported in ref.\ \cite{StenhammarTAMC2013}, but found there to be reproducible in an integrable Cahn-Hilliard model, albeit with a non-polynomial $f_{0}$, for which the asymptotic $1/3$ power is assured. This suggests a crossover scenario in which all gradient terms, whether active or passive, gradually merge into an effective interfacial tension that drives $t^{1/3}$ coarsening, once $L$ is very much larger than the interfacial width. Further arguments for this outcome are given below and in Supplementary Notes 2 and 3. While the case for a standard asymptotic $t^{1/3}$ scaling law is compelling, our next result shows that the $\lambda$ term in Active Model B is certainly not representable \textit{solely} by a shift in interfacial tension, since this would have no effect on phase equilibria.

\subsection{Failure of the common tangent construction}
In Passive Model B ($\lambda=0$), bulk phase separation is characterized by two static coexisting phases of infinite extent that are separated by a planar interface. These bulk phases have densities $\phi\to-1$ and $\phi\to 1$ for $\lvert z\rvert\to\infty$, respectively, where $z$ is a coordinate perpendicular to the interface. 
The densities of the coexisting phases are set by the common tangent construction on $f_{0}(\phi)$, as is fully explained below, and imply $\mu_{0}=0$ for $\lvert z\rvert\to\infty$.
Furthermore, the constant densities for $\lvert z\rvert\to\infty$ imply $\mu_{1}=0$ and therefore $\mu=0$ in both bulk phases. 

At first sight one expects the common tangent construction still to be possible for finite $\lambda$, because the active contribution to the nonequilibrium chemical potential, $\mu^{\mathrm{A}}_{1}$, vanishes in both bulk phases, and the construction itself cares only about the bulk free-energy density $f_{0}(\phi)$ and not about interfacial tension. This view, however, is mistaken because the construction implicitly assumes integrability, i.e., the existence of a free-energy structure everywhere, including any gradient terms. Indeed we prove here that no similar solution exists at $\mu = 0$ for Active Model B,  which means that the common tangent construction fails whenever $\lambda\neq 0$.

Instead, for $\mu_{0}=0$ we find only solutions describing spatially periodic (lamellar) phases. At first sight, these findings suggest that micro-phase separation, rather than coexisting bulk phases, might be the generic fate of Active Model B. This is, for example, what happens when Passive Model B is coupled to a somewhat different form of activity, namely logistic population growth \cite{CatesMPT2010}. It is also hinted at by various experiments in which active particles form clusters whose size seemingly remains finite at long times \cite{PalacciSSPC2013,TheurkauffCBPYB2012,SchwarzLinekVCCMP2012}.
Such an outcome would however be paradoxical given our numerical finding of bulk demixing rather than micro-phase separation in Active Model B. To resolve this paradox,we show below that a planar interface \textit{does} exist between fully phase-separated states; but their bulk densities are \textit{not} given by a common tangent construction on $f_{0}$.

To prove this, we look for a fully phase-separated state with a planar interface and take advantage of its two translational invariances to reduce the problem to one spatial dimension. 
For the required static solution of Active Model B this means $\phi = \phi(z)$ where $z$ is a coordinate normal to the interface. To describe static bulk phase separation, the current in eq.\ \eqref{J} (with $\mathbfup{\Lambda}=\mathbfup{0}$), specifically its $z$ component $J_{z}$, must vanish. 
(We assume a uniform current is excluded by boundary conditions at infinity.)
This requires $\mu$ to be constant. Let us first try to find its value by following the usual equilibrium reasoning. Constant $\mu$ requires equality of the bulk chemical potential terms $\mu_{0} = \dif f_{0}/\dif\phi$
and also of the bulk thermodynamic pressures $P = \phi\mu_{0}-f_{0}$ in the two phases. 
(Recall that in nonequilibrium systems there is no general relation between the thermodynamic pressure thus defined and the mechanical pressure; we return to this point below.) 
Together, these conditions imply that a common tangent can be drawn to the curve $f_0(\phi)$ at the two coexisting densities. Since $f_0$ is symmetric, this implies $\mu\to 0$ for $\lvert z\rvert\to\infty$, so that by this argument the nonequilibrium chemical potential $\mu$, which differs from $\mu_0$ only in interfacial regions, is zero everywhere.  
The static density profile $\phi(z)$ is thus given by  
\begin{equation}
-\phi+\phi^{3}-\phi''+\lambda(\phi')^{2}=0 \;.
\label{SI:eq:phiBPS}%
\end{equation}
By renaming $z$ and $\phi$ according to ($z\to t$, $\phi\to x$), this equation for the density profile $\phi(z)$ can be mapped onto the equation of motion for the trajectory $x(t)$ of a Newtonian particle of unit mass in a symmetric inverted potential $U(x)=-f_{0}(x)$ under the influence of a velocity-dependent force $\lambda\dot{x}^{2}$ that is invariant under time reversal.
The resulting equation of motion is 
\begin{equation}
\ddot{x}=-U'(x)+\lambda\dot{x}^{2} \;. 
\label{SI:eq:xBPS}%
\end{equation}
This makes calculating the density profile $\phi(z)$ the same problem as finding the trajectory $x(t)$ of a Newtonian particle in the potential $U(x)$ with a velocity-dependent force arising from the $\lambda$ term. In the passive case ($\lambda =0$), this \textit{Newton mapping} is a standard procedure (see ref.\ \cite{OxtobyH1982} and references therein), whose details we recall in the Methods section. In brief, the domain-wall solution is described by a particle that leaves one of the two maxima in $U(x)$ with infinitesimal velocity, travels across the valley in the inverted potential, and then comes to rest at the other maximum. 

Given this picture, let us consider the effect of $\lambda\neq 0$, i.e., Active Model B. This creates a velocity-dependent force in eq.\ \eqref{SI:eq:xBPS}, $\lambda\dot{x}^{2}$, which has the same sign throughout the trajectory. 
Because of this, if the Newtonian particle starts very near the top of the first maximum of $U(x)$, i.e., if one starts from a large bulk domain close to the bulk density $\phi_{1}=\pm 1$, the particle either does not make the top of the other maximum or it overshoots.
The former gives periodic oscillations between the initial coordinate $x_{1}$ and the turning point $x_{2}$, which physically describe micro-phase separation, while the latter gives an unphysical blowup. 
In the case of micro-phase separation the Newtonian particle retraces its steps from the turning point $x_{2}$, which is finitely below the second peak, and regains exactly the kinetic energy it lost to the velocity-dependent force during the first half of the cycle. So it arrives exactly at $x_{1}$ where it started, comes virtually to rest there, creating another large domain of $\phi_{1}$, before turning round and starting the cycle again. 
During this periodic motion, the particle only briefly visits the even peaks, creating a micro-domain of the $\phi_{2}$ phase whose width is of order of the interfacial width.
To make the domain of the second phase any wider, the velocity-dependent term has to be made exponentially small (i.e., these domains have a width that varies as $-\log|\lambda|$).
Only this will allow the particle to closely approach the top of the other maximum. 

If the Newtonian particle, on the other hand, starts very near the top of the second maximum of $U(x)$, i.e., if one starts with a very large domain of the second phase, and $\lambda$ has the sign that prevents the particle from overshooting when it starts from the first maximum, the $\lambda$ term gives energy to the particle so that it now overshoots. 
These arguments confirm that for $\lambda\neq 0$ and $\mu=0$, static bulk phase separation is impossible. 

Three resolutions can be envisaged. Either there ceases to be a static solution of Active Model B, or the solutions describing micro-phase separation become stable, or $\mu$ is nonzero and the common tangent construction fails. 
We show next that the last of these resolutions applies.

\subsection{The uncommon tangent construction}
We now generalize the discussion from the previous section to allow a nonzero nonequilibrium chemical potential $\mu\neq 0$ obeying eq. \eqref{mu}. We show that for Active Model B a solution with a planar interface between two static bulk coexisting phases does after all exist, but is shifted to a nonzero $\mu = \mu_{\mathrm{b}}$ so that the coexisting densities are not $\phi=\pm 1$. We also derive the resulting nonequilibrium chemical potential at bulk phase coexistence $\mu_{\mathrm{b}}$ as a function of the parameter $\lambda$.

\begin{figure}
\centering
\includegraphics[width=\columnwidth]{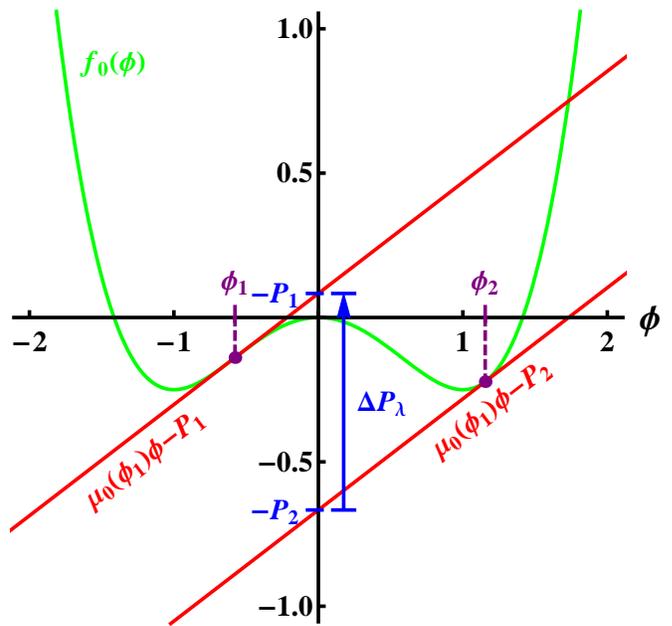}
\caption{\label{Fig.4}\AU{`Uncommon tangent' construction.} $\mu$ is equal in coexisting phases and the thermodynamic pressures $P_{1}$ and $P_{2}$ in each phase are the negative intercepts on the vertical axis. Here $\phi_{1}$ and $\phi_{2}$ are the densities of the two coexisting phases and $\Delta P_{\lambda}=P_{2}-P_{1}$ is the pressure difference. This plot shows the construction for $\lambda\to\infty$, where $\mu_{0}(\phi_{1})=\mu_{\mathrm{c}}$.}
\end{figure}

For $\mu\neq 0$ eq.\ \eqref{SI:eq:phiBPS} for the density profile $\phi(z)$ becomes 
\begin{equation}
-\phi+\phi^{3}-\phi''+\lambda(\phi')^{2}=\mu 
\label{SI:eq:muID}%
\end{equation}
and the corresponding Newtonian equation is eq.\ \eqref{SI:eq:xBPS} with the now asymmetric potential $U(x)=\mu x -f_{0}(x)$.
We define a critical value of $|\mu| = \mu_{\mathrm{c}}$ such that, whenever $\lvert\mu\rvert<\mu_{\mathrm{c}}$, the potential $U(x)$ has two maxima at positions $x^{(1)}_{\mathrm{max}}$ and $x^{(2)}_{\mathrm{max}}$, with $x^{(1)}_{\mathrm{max}}<x^{(2)}_{\mathrm{max}}$, and a minimum between them.
When $\mu\neq 0$ the heights of the two maxima of $U(x)$ are different, and the sign of $\mu$ decides which maximum is lower.
At $|\mu|=\mu_{\mathrm{c}}$ the minimum and the lower maximum merge into a saddle point, and as $|\mu|$ increases beyond the critical value $\mu_{\mathrm{c}}$, there remains only a single maximum. (We find $\mu_{\mathrm{c}}=3^{-1/2}-3^{-3/2}\approx 0.38$.) 
Notice that $\mu_{\mathrm{c}}$ can also be defined as the maximum slope with which two distinct parallel tangents can be drawn on $f_{0}(\phi)$ (see Fig.\ \ref{Fig.4}). 
This maximum slope arises when one such tangent passes through the inflection point of $f_{0}(\phi)$, also known as the spinodal point.

Due to these properties of the potential $U(x)=\mu x -f_{0}(x)$, for a Newtonian particle with the equation of motion \eqref{SI:eq:xBPS} there are three different types of motion possible: i) a periodic oscillatory motion between two turning points $x_{1}$ and $x_{2}$ with $x^{(1)}_{\mathrm{max}}<x_{1}<x_{2}<x^{(2)}_{\mathrm{max}}$ describing micro-phase separation, ii) the non-periodic limiting case $x_{1}=x^{(1)}_{\mathrm{max}}$ and $x_{2}=x^{(2)}_{\mathrm{max}}$ that corresponds to bulk phase separation, and iii) divergent solutions corresponding to unphysical density profiles. 
Since there is only one maximum for $\lvert\mu\rvert\geqslant\mu_{\mathrm{c}}$, micro-phase separation and bulk phase separation are possible only for $\lvert\mu\rvert<\mu_{\mathrm{c}}$. 
For $\lvert\mu\rvert\geqslant\mu_{\mathrm{c}}$, on the other hand, no physical solution for the static density profile $\phi(z)$ exists. 

\begin{figure}
\centering
\includegraphics[width=\columnwidth]{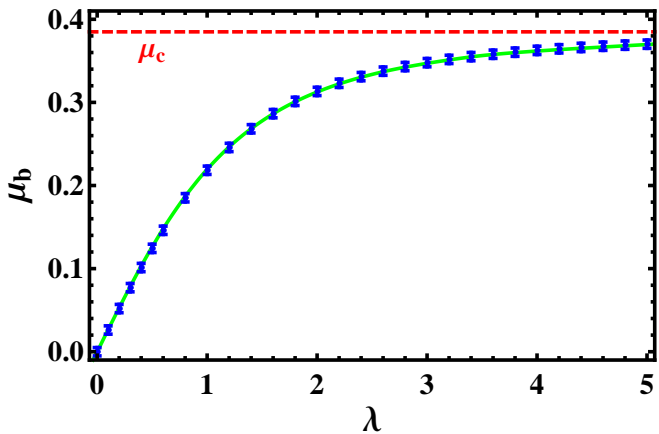}
\caption{\label{Fig.3}\AU{Nonequilibrium chemical potential at bulk phase coexistence.} Analytical result (solid line) and numerical results obtained from simulations using Active Model B (points with error bars that denote the estimated discretization error) for $\mu_{\mathrm{b}}(\lambda)$. For $\lambda\to\infty$ the function $\mu_{\mathrm{b}}(\lambda)$ asymptotically tends to $\mu_{\mathrm{c}}=3^{-1/2}-3^{-3/2}$ (dashed line). Only positive $\lambda$ is shown, since $\mu_{\mathrm{b}}(\lambda)$ has the symmetry property $\mu_{\mathrm{b}}(-\lambda)=-\mu_{\mathrm{b}}(\lambda)$.}
\end{figure}

In the following we consider the situation $\lvert\mu\rvert<\mu_{\mathrm{c}}$ and treat the appearance of bulk phase separation as a special limiting case of micro-phase separation. 
The density distribution $\phi(z)$ for bulk phase separation is a special solution of eq.\ \eqref{SI:eq:muID} or equivalently eq.\ \eqref{SI:eq:xBPS} with $U(x)=\mu x -f_{0}(x)$ for an appropriate value  $\mu =\mu_{\mathrm{b}}$ of the nonequilibrium chemical potential, which depends on the parameter $\lambda$.
Equation \eqref{SI:eq:xBPS} with $U(x)=\mu x -f_{0}(x)$ is an autonomous nonlinear second-order ordinary differential equation (ODE) that cannot be solved analytically. 
The function $\mu_{\mathrm{b}}(\lambda)$, however, can be derived from this equation even without solving it. The procedure is detailed in the Methods section and gives an analytic result of $4/15$ for the slope of $\mu_{\mathrm{b}}(\lambda)$ at $\lambda=0$ as well as an implicit function for $\mu_{\mathrm{b}}(\lambda)$, which is plotted in Fig.\ \ref{Fig.3}. 

With a much higher computational expense the function $\mu_{\mathrm{b}}(\lambda)$ can also be determined from simulations
that solve the dynamical equation \eqref{phidot} of Active Model B by evolving until a steady-state interfacial profile is reached. The results of such simulations are also shown in Fig.\ \ref{Fig.3}.   
The function $\mu_{\mathrm{b}}(\lambda)$ is zero for $\lambda=0$, increases monotonically for $\lambda>0$, and asymptotically approaches the limiting critical value $\mu_{\mathrm{b}}(\infty)=\mu_{\mathrm{c}}$ for $\lambda\to\infty$. 
Using the function $\mu_{\mathrm{b}}(\lambda)$, the coexisting densities $\phi_{1}=\phi(-\infty)$ and $\phi_{2}=\phi(\infty)$ are given by the smallest and largest solution of $U'(\phi)=0$.  

The above arguments demonstrate that solutions describing static bulk phase separation exist for all values of $\lambda$, 
but are shifted to $\mu=\mu_{\mathrm{b}}\neq 0$ and involve coexisting densities $\phi\neq\pm 1$. Other static solutions describing, e.g., lamellar micro-phase separation, also exist but are generally not the result of numerically solving eq.\ \eqref{phidot}. Such states are generically unstable to the same Ostwald ripening mechanism as in the passive case (see Methods section); this is confirmed in Supplementary Note 3 for the case of droplets in higher dimensions.  

Bulk phase coexistence in Active Model B is thus governed by an `uncommon tangent' construction in which tangents to $f_{0}(\phi)$ are parallel, but displaced from each other (see Fig.\ \ref{Fig.4}). 
Defining as usual the bulk thermodynamic pressure by $P=\phi\mu_{0}-f_{0}$, the $\lambda$-induced vertical displacement is identified as a pressure difference $\Delta P_{\lambda}$ between the coexisting bulk phases. The common tangent construction equates both $\mu_{0}$ and $P$ in these phases; its failure is thus attributable to a $\lambda$-induced \textit{jump in thermodynamic pressure} across the planar interface that separates them. 
This is a very interesting, and somewhat unexpected, consequence of activity and the resulting breakdown of detailed balance in the interfacial regions.

\subsection{Active pressure competes with Laplace pressure}
The active pressure jump $\Delta P_{\lambda}$ has no direct counterpart in passive systems. However, for a passive droplet with radius $R$ the interfacial tension $\gamma$ (where $\gamma=\sqrt{8}/3$ in Passive Model B \cite{KendonCPDB2001}) creates a Laplace pressure jump $\Delta P_{\mathrm{L}} = (d-1)\gamma/R$ across the interface. In Supplementary Note 2, we establish that to leading order in large $R$ (i.e., small $\lvert\lambda\rvert$), a static $\mu=0$ solution exists for a droplet of radius $R^{*}=(5/\sqrt{8})(d-1)/\lvert\lambda\rvert$ at which the active and Laplace pressures are equal and opposite; common tangency is thereby restored. This can happen only for one of the two possible dispositions of internal and external phases, set by the sign of $\lambda$. This result shows that the activity-induced pressure jump $\Delta P_{\lambda}$ has, within our mapping from active onto passive phase separation, the same `thermodynamic' status as the Laplace pressure. 

The existence of these static droplet solutions raises once again the possibility of micro-phase separation. For instance one could envisage a state of droplets, each with radius $R^{*}$, embedded in a continuous phase at zero $\mu$.
If stable, such a phase might explain various experiments showing formation of small finite clusters in bacteria and artificial active colloidal particles \cite{SchwarzLinekVCCMP2012,TheurkauffCBPYB2012,PalacciSSPC2013}. (On dimensional grounds, $R^{*}\simeq v\tau$.) However, this would contradict the numerics reported above. We resolve this by noting that even the single droplet solution is unstable, as shown in Supplementary Note 2. The instability appears to be fundamentally no different from the classical one of a finite fluid droplet in unstable equilibrium with its vapour. Indeed, in Passive Model B a stationary value of the droplet radius $R_{0}$ exists for $\mu>0$; but this is an unstable fixed point separating shrinking from growing droplet states. 
Moreover, as shown quantitatively in Supplementary Note 3, the active pressure $\Delta P_{\lambda}$ offsets the relation between $\mu$ and the droplet size, but cannot halt the coarsening of an assembly of droplets by Ostwald ripening, in which small ones evaporate and large ones grow. This reasoning again points firmly towards the usual diffusive growth law $L\sim t^{1/3}$, which holds for Ostwald-like dynamics in both droplet and bicontinuous morphologies \cite{Bray1994}, and explains in large part the remarkable similarity between active and passive phase separation. It also explains why, in the active case, micro-phase separation is \textit{not} the generic end-point of the dynamics; via Ostwald ripening, coarsening proceeds indefinitely. The deviations from $\alpha = 1/3$ found numerically are, in this view, almost certainly transient or crossover effects.

\subsection{Mechanical versus thermodynamic pressure}
In future work we will generalize this study to allow for coupling of an active scalar field to a momentum-conserving solvent flow. There we will present an active counterpart of `Model H' which describes such coupling for passive systems \cite{HohenbergH1977,ChaikinL1995}. This raises an intriguing issue concerning the nature of the pressure, which we have defined in this paper as $P = \phi\mu_{0}-f$ to coincide with the standard bulk thermodynamic definition. In passive systems, this is of course equal to the mechanical pressure defined either as a force per unit area on a boundary, or via the diagonal part of the stress tensor. Accordingly for instance, the Laplace pressure jump across a curved interface can be measured directly by a mechanical probe in a passive system.

However, the equivalence of mechanical and thermodynamic pressure in such a system stems ultimately from the fact that the same interparticle forces determine both mechanics and thermodynamics. In contrast, for active particles, even the integrable part of the free energy has no simple link to interparticle forces: instead it encodes the effects of density on self-propulsion through the mapping of ref.\ \cite{TailleurC2008}. This means that the quantity $P$ found via that mapping cannot generally be viewed as an actual mechanical pressure. 
This is unsurprising for a system far from equilibrium; indeed mechanical and thermodynamic pressures differ even in systems quite close to equilibrium, such as flowing fluids.  
The active pressure jump across a flat interface could not therefore be measured with a pressure gauge, but is still a pressure jump in the thermodynamic sense that $P=\phi\mu-f$ has different values in the two bulk phases. For a curved interface, exactly the same remarks apply to the Laplace pressure contribution, whenever the phase separation itself, and hence the resulting interfacial tension, is activity-driven. Indeed the activity-driven interfacial tension is itself not detectable with a tensiometer; its meaning stems from its ability to drive a diffusive flux of active particles via the nonequilibrium chemical potential appearing in eq.\ \eqref{J}.   

When coupling the active system to a momentum-conserving solvent via the Navier-Stokes equation, it is the mechanical pressure that enters, not the thermodynamic one. As we will address in detail elsewhere, in practice this means that an Active Model H must combine the diffusive dynamics of Active Model B developed here with a conceptually separate account of the mechanical forces created by self-propulsion. 

That said, if we restrict attention to the class of systems where bulk phase separation is caused solely by interparticle attractions, so that the only effect of weak activity is to create a nonzero value of $\lambda$, even the mechanical pressure in the two bulk phases is then unequal. This is because the gradient terms controlled by $\lambda$  vanish in bulk: therefore ordinary equilibrium thermodynamics holds locally, and mechanical and thermodynamic pressure must once again coincide, in each bulk phase. The activity represented by $\lambda$ plays a direct role only near the interface, where it creates the uncommon tangent condition. This obliges the system to develop a real, physically measurable mechanical pressure jump between phases, equal to our active pressure.
Note that our active pressure is not just the mechanical pressure that a confined active gas exerts on a container wall \cite{MallorySVC2014} and that results directly from the increased speed and surface accumulation of active particles compared to passive ones. In contrast to the pressure of an active gas our pressure jump across the interface between a dilute and a dense phase of active particles is not proportional to the density and its origin is more subtle.

\section{Discussion}%
We have argued that, when no solvent is present so that dynamics is diffusive, the physics of active-particle phase separation is fully captured by Active Model B. This combines a $\phi^{4}$ bulk free energy with passive and active gradient terms in a minimal fashion. It represents an intriguing class of problems involving diffusive phase separation in systems where detailed-balance violations are created primarily by interfaces. The minimalist structure of Active Model B allows not only for efficient simulation, but also for several analytic results to be obtained. These explain why such detailed-balance violations have paradoxically small effects on coarsening dynamics, for which interfacial physics is usually dominant, but large ones on the phase diagram, for which such physics is, at first sight, irrelevant. Somewhat similar equations have been used recently to study crystal growth at finite undercooling \cite{WatsonN2006,WatsonORD2003} and state selection in shear-banding rheology \cite{Olmsted2008}. In these cases it is known that nonintegrable gradient terms can destroy the common tangent construction for phase coexistence \cite{WatsonORD2003,Olmsted2008}. In Active Model B the physics that replaces common tangency is both simple and remarkable: the equality of chemical potential between phases is maintained, but activity creates a direct analog of the thermodynamic Laplace pressure operating across the interface between bulk phases which, unlike its equilibrium counterpart, remains finite even for a planar interface.

Our work sheds direct light on how the nonintegrable term in Active Model B leads to new and unintuitive physical predictions for active-particle phase separation. The role played by this active term can best be understood in terms of 
the well-established bulk mapping between active and passive phase 
separation, as extended here to include an interfacial tension plus an 
interfacial pressure jump that has no passive counterpart.
Our results are all the more powerful and surprising because we have so far found no convincing route to obtain them by qualitative reasoning applied directly to the motion of active particles. It is very difficult for such reasoning to capture the unusual structure of the problem, which involves restoration of TRS (absent microscopically) at zeroth order in spatial gradients and its loss again at higher order. A fully microscopic interpretation of our results may therefore remain elusive.

\section{Methods}%
\subsection{Numerical analysis}
In order to solve Active Model B numerically, a finite-difference scheme with periodic boundary conditions was applied. 
The initial spatial distribution of the order-parameter field $\phi$ was random for the numerical calculations whose results are shown in Figs.\ \ref{Fig.1} and \ref{Fig.2} and a step function (with $\phi=-1$ for $64\leqslant z\leqslant 192$ and $\phi=1$ otherwise) for the results shown in Fig.\ \ref{Fig.3}.
Both for simulations in two and three spatial dimensions, the time step size was $\Delta t=0.001$, while the spatial step size was $\Delta z=0.5$ for two dimensions and $\Delta z=1$ for three dimensions. 
The lattice size was mainly $256\times 256$ for two dimensions with the exception $256\times 50$ for the numerical calculations corresponding to Fig.\ \ref{Fig.3} and $128\times 128\times 128$ for three dimensions. 
Finally, the domain length scale $L(t)$ in Fig.\ \ref{Fig.2} was calculated from the inverse of the
first moment of the spherically averaged structure factor $S(k,t)$ \cite{KendonCPDB2001}:
\begin{equation}
L(t)=2\pi\frac{\int\:\!\! S(k,t)\,\dif k}{\int\:\!\! kS(k,t)\,\dif k} \;.
\end{equation}
The spherically averaged structure factor is defined as 
\begin{equation}
S(k,t)=\langle\phi(\mathbfup{k},t)\phi(-\mathbfup{k},t)\rangle_{k} \;,
\end{equation}
where $k=\lvert\lvert\mathbfup{k}\rvert\rvert$ is the modulus of the wave vector $\mathbfup{k}$, 
$\phi(\mathbfup{k},t)$ is the spatial Fourier transform of the order-parameter field $\phi(\mathbfup{r},t)$,
and $\langle\,\cdot\,\rangle_{k}$ denotes an average over a shell in $\mathbfup{k}$ space at fixed $k$.

\begin{figure*}
\centering
\includegraphics[width=\textwidth]{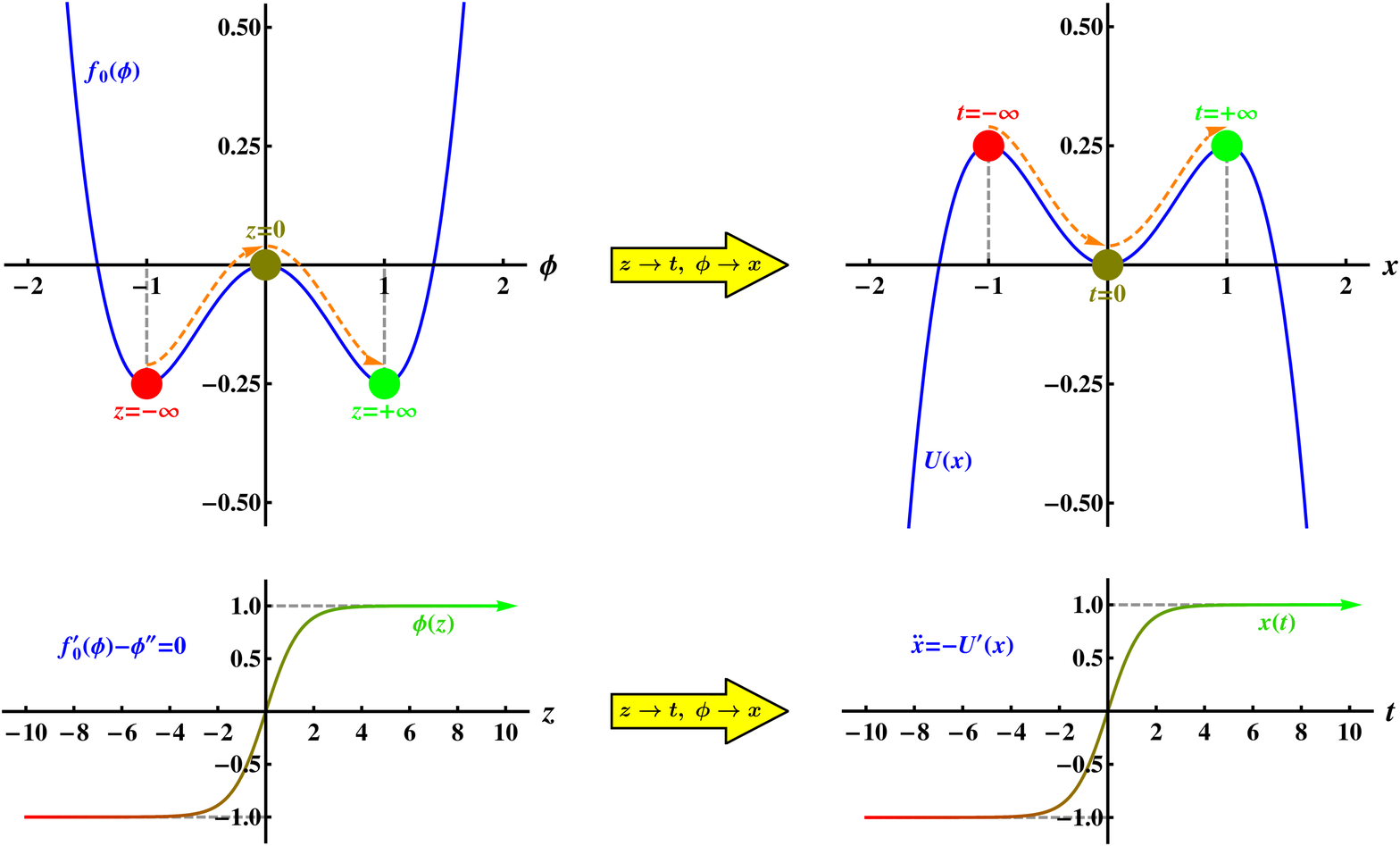}
\caption{\label{Fig.5}\AU{Sketch of the Newton mapping.} Calculating the density profile $\phi(z)$ of a system with free-energy density $f_{0}(\phi)$ is the same problem as finding the trajectory $x(t)$ of a Newtonian particle in the external potential $U(x)=-f_{0}(x)$. This sketch shows bulk phase separation in Passive Model B for $\lambda=0$ and $d=1$ as an example.}
\end{figure*}

\subsection{Newton mapping}
Here we summarize the Newton mapping for the passive case where $\lambda = 0$ (see Fig.\ \ref{Fig.5} for a sketch).  
The inverted potential $U(x) = -f_0(x)$ has two maxima of equal height at $x=\pm 1$ with a minimum in between. 
For $x<-1$ and $x>1$ it falls off to minus infinity.
Notice that the starting position $x_{1}=x(-\infty)$ of the Newtonian particle with the equation of motion \eqref{SI:eq:xBPS} corresponds to the density $\phi_{1}=\phi(-\infty)$ of the first bulk phase. 
The Newtonian particle can start from rest from one of three qualitatively different domains: i) $-1<x_{1}<1$, ii) $x_{1}\in\{-1,1\}$, and iii) $x_{1}<-1$ or $x_{1}>1$.
In case i) the Newtonian particle oscillates indefinitely between its starting position $x_{1}$ and a turning point at position $-x_{1}$. For the density distribution $\phi(z)$ this describes a lamellar state corresponding to micro-phase separation. (This state could be viewed in one dimension as a series of droplets each of exactly equal size and separation.) Such a state is unstable (in any dimension) to small density variations causing large droplets to grow at the expense of small ones. This is the Ostwald process discussed further in Supplementary Note 3.

On the other hand, in the more particular case ii) the Newtonian particle starts from rest infinitesimally below one peak of the potential $U(x)$, where it hovers a long time before starting to move, shoots up the other side and then comes to rest infinitesimally below the other peak.
This can be done in either direction and gives the familiar domain-wall solution that describes bulk phase separation.
Finally, in case iii) the Newtonian particle accelerates indefinitely towards $x=-\infty$ or $x=\infty$. This corresponds to a divergent and unphysical solution of eq.\ \eqref{SI:eq:phiBPS}, which is anyway ruled out by the boundary conditions. Imposing $\phi(z) = \pm 1$ at $\lvert z\rvert\to\infty$ selects the bulk interfacial profile as the only stable, nonuniform, static solution of Passive Model B.

\subsection{Determination of the coexistence condition}
To find the function $\mu_{\mathrm{b}}(\lambda)$ we first reduce the order of eq.\ \eqref{SI:eq:xBPS} with $U(x)=\mu x -f_{0}(x)$. The substitution $\nu(x)=\dot{x}(t)$ transforms it into a non-autonomous nonlinear first-order ODE.      
A further non-invertible substitution $w(x)=\nu^{2}(x)$ then leads to the linear first-order ODE
\begin{equation}
w'(x)=-2 U'(x)+2\lambda w(x) \;.
\label{SI:eq:w}%
\end{equation}
This equation describes the squared velocity $w(x)$ of a Newtonian particle, with the equation of motion \eqref{SI:eq:xBPS}, as a function of the particle's position $x$.
Equation \eqref{SI:eq:w} is much simpler than the ODE \eqref{SI:eq:xBPS} with $U(x)=\mu x -f_{0}(x)$ and can be solved analytically as 
\begin{equation}
w(x)=w_{0} e^{2\lambda(x-x_{0})}-\frac{1}{8\lambda^{4}}\big(g_{\lambda}(x)-g_{\lambda}(x_{0})e^{2\lambda(x-x_{0})}\big)
\label{SI:eq:w_Lsg}%
\end{equation}
with the initial value $w_{0}=w(x_{0})$ at the initial position $x_{0}=x(0)$, and the polynomial
$g_{\lambda}(x)=6+12\lambda x+4\lambda^{2}(3x^{2}-1)+8\lambda^{3}(-\mu-x+x^{3})$.
 
With the help of eq.\ \eqref{SI:eq:w_Lsg} the condition for micro-phase separation can be written as follows: there are two positions $x_{1}$ and $x_{2}$ with $x^{(1)}_{\mathrm{max}}<x_{1}<x_{2}<x^{(2)}_{\mathrm{max}}$ so that $w(x_{1})=w(x_{2})=0$. 
This condition is equivalent to $h_{\lambda}(x_{1})=h_{\lambda}(x_{2})$ with $h_{\lambda}(x)=g_{\lambda}(x)e^{-2\lambda x}$.  
Bulk phase separation as a limiting case of micro-phase separation appears for $x_{1}=x^{(1)}_{\mathrm{max}}$ and $x_{2}=x^{(2)}_{\mathrm{max}}$. 
The condition for bulk phase separation can thus be written as 
\begin{equation}
h_{\lambda}(x^{(1)}_{\mathrm{max}})=h_{\lambda}(x^{(2)}_{\mathrm{max}}) \;.
\label{SI:eq:hps}%
\end{equation}
Notice that the last term in $g_{\lambda}(x)$ and therefore also in $h_{\lambda}(x)$ vanishes for $x\in\{x^{(1)}_{\mathrm{max}},x^{(2)}_{\mathrm{max}}\}$.
The condition \eqref{SI:eq:hps} depends only on $\mu$, which we denote as $\mu_{\mathrm{b}}$ in the case of bulk phase separation, and $\lambda$.
This means that bulk phase separation requires a relation between $\mu_{\mathrm{b}}$ and $\lambda$ described by a function $\mu_{\mathrm{b}}(\lambda)$.

For $\lvert\lambda\rvert \ll 1$ (which also implies $\lvert\mu_{\mathrm{b}}\rvert\ll 1$; notice that $\mu_{\mathrm{b}}=0$ for $\lambda=0$) the function $\mu_{\mathrm{b}}(\lambda)$ can be found from the implicit equation \eqref{SI:eq:hps} by expanding the latter in both $\lambda$ and $\mu_{\mathrm{b}}$. The resulting perturbative solution of eq.\ \eqref{SI:eq:hps} is given by 
\begin{equation}
\mu_{\mathrm{b}}(\lambda)=\frac{4}{15}\lambda + \mathcal{O}(\lambda^{3}) \;.
\label{SI:eq:mub}%
\end{equation}
Also the property $\mu_{\mathrm{b}}(-\lambda)=-\mu_{\mathrm{b}}(\lambda)$ follows from eq.\ \eqref{SI:eq:hps}, so that there are no terms of even order in eq.\ \eqref{SI:eq:mub}.

\bibliographystyle{naturemag}
\bibliography{refs}

\section{Acknowledgements}
Seminal discussions with Ronojoy Adhikari, Sriram Ramaswamy, Julien Tailleur, and Stephen J.\ Watson are acknowledged. We thank EPSRC EP/J007404 for funding. R.W.\ gratefully acknowledges financial support through a Postdoctoral Research Fellowship (WI 4170/1-1) from the German Research Foundation (DFG), J.S.\ gratefully acknowledges financial support from the Swedish Research Council (350-2012-274), R.J.A.\ through a Royal Society University Research Fellowship, and M.E.C.\ through a Royal Society Research Professorship. M.E.C.\ thanks the Isaac Newton Institute (Cambridge, UK) for hospitality.

\section{Author contributions}
All authors designed and performed the research and analysed the data; 
R.W. and A.T. prepared the figures; 
R.W. and M.E.C. wrote the paper with input from the other authors.

\appendix
\clearpage

\section{Supplementary Note 1: Derivation of Active Model B}%
We start from a specific model of microscopic dynamics (with discrete or continuous
angular reorientation) comprising particles whose propulsion speed $v(\rho)$ is a decreasing function of the particle number density $\rho$ \cite{TailleurC2008,CatesT2013}. 
The linear transformation between $\rho$ and the order-parameter field $\phi$ is given by
\begin{equation}
\rho = \bar{\rho}+\phi\:\!\frac{\rho_{\mathrm{H}}-\rho_{\mathrm{L}}}{2} \;, 
\label{SI:eq:rho}
\end{equation}
where $\bar{\rho}=(\rho_{\mathrm{H}}+\rho_{\mathrm{L}})/2$, and
$\rho_{\mathrm{L}}$ and $\rho_{\mathrm{H}}$ are the low and high values of $\rho$ as found from a common tangent construction on the free-energy density \cite{TailleurC2008}
\begin{equation}
f_{\rho}=\rho(\ln(\rho)-1)+\int^{\rho}_{0} \!\!\!\ln(v(u))\,\dif u \;.
\label{SI:eq:frho}%
\end{equation}%
As derived in ref.\ \cite{StenhammarTAMC2013}, the leading-order gradient term of the nonequilibrium chemical potential $\mu$ reads
\begin{equation}
\mu_{1} = -\kappa(\phi)\nabla^{2}\phi \;.
\label{SI:munab}%
\end{equation}
Here $\kappa(\phi)\propto -v(\phi) v'(\phi)$ is a positive but non-constant quantity, whose form is directly computable from the density-dependent swim speed $v(\rho)$. We may now split $\mu_{1} = \mu^{\mathrm{P}}_{1}+\mu^{\mathrm{A}}_{1}$ into two parts. The first is an integrable `passive' term
\begin{equation} 
\begin{split}%
\mu^{\mathrm{P}}_{1}&= \frac{\delta}{\delta\phi}\int\! f^{\mathrm{P}}_{1}  \,\dif^{d}r 
= -\tilde{\kappa}(\phi)\nabla^{2}\phi -\frac{\tilde{\kappa}'(\phi)}{2}(\nabla\phi)^{2} 
\end{split}%
\label{SI:muP1}%
\end{equation}%
with the free-energy density $f^{\mathrm{P}}_{1} =\tilde\kappa(\phi)(\nabla\phi)^{2}/2$ and a function $\tilde{\kappa}(\phi)$ chosen below. The second is an inherently `active' term, $\mu^{\mathrm{A}}_{1}$, that is not a functional derivative and so nonintegrable. 

In principle there are many ways to do this splitting involving different choices for $\tilde{\kappa}(\phi)$. For instance one might simply assert that eq.\ \eqref{SI:munab} is not a functional derivative and on this basis say that this is the active gradient term $\mu^{\mathrm{A}}_{1}$ and there is no passive one $\mu^{\mathrm{P}}_{1}$. This would, however, falsely classify as active the fully integrable and hence (in this language) passive case of constant $\kappa$.
This pitfall is avoided by choosing $\tilde{\kappa}(\phi) = \kappa(\phi)$, so that the $\nabla^{2}\phi$ term in $\mu_{1}$ is assigned wholly to the passive sector, but is then accompanied by an additional passive term proportional to $\kappa'(\phi)(\nabla\phi)^{2}$.
The absence of this second term in eq.\ \eqref{SI:munab} implies that the active part can now be identified as $\mu^{\mathrm{A}}_{1} = \kappa'(\phi)(\nabla\phi)^{2}/2$.
This motivates the choice $\mu^{\mathrm{A}}_{1} = \lambda(\nabla\phi)^{2}$ made in eq.\ \eqref{mu} for Active Model B, with $\lambda$ now a constant coefficient of the same order of magnitude as $\kappa'(\phi)$. Constancy of $\lambda$ is clearly the minimal nontrivial choice for $\mu^{\mathrm{A}}_{1}$.  

The above considerations establish $\mu^{\mathrm{A}}_{1} = \lambda(\nabla\phi)^{2}$ as the simplest nontrivial form for the nonintegrable gradient contributions in active-particle systems. We can revert to the simplest nontrivial form for the remaining (integrable) gradient terms also. That is, we can follow standard practice with Passive Model B: in the integrable square-gradient free-energy density $f^{\mathrm{P}}_{1} = \kappa(\phi)(\nabla\phi)^{2}/2$ we first suppress the $\phi$ dependence of $\kappa$ and then choose units where $\kappa=1$.
(In these units $\lambda$, while negative, is of order unity.) On this basis we set $\mu^{\mathrm{P}}_{1} = -\nabla^{2}\phi$ in eq.\ \eqref{mu}, which completes the specification of Active Model B as defined in the main text.

Explicit coarse-graining of the dynamics of active Brownian particles (ABPs) leads to the gradient term \cite{StenhammarTAMC2013}
\begin{equation}
\mu^{\rho}_{1} = -\kappa_{\rho}(\rho)\nabla^{2}\rho 
\label{SI:munabrho}%
\end{equation}
with the function $\kappa_{\rho}(\rho)=-\gamma^{2}_{0}\tau^{2}v(\rho)v'(\rho)$, where the order-unity parameter $\gamma_{0}$ defined in ref.\ \cite{StenhammarTAMC2013} relates the nonlocality of $v(\rho)$ to the persistence length $v(\rho)\tau$. 
For ABPs the density-dependent swim speed $v(\rho)$ is given by $v(\rho)=v(0)(1-v(0)\sigma_{\mathrm{s}}\tau_{\mathrm{c}}\rho)$ with a scattering cross section $\sigma_{\mathrm{s}}$ and the duration $\tau_{\mathrm{c}}$ a particle is effectively stalled at a collision event \cite{StenhammarTAMC2013}. 
Using eq.\ \eqref{SI:eq:rho} to express the particle number density $\rho$ in terms of the order-parameter field $\phi$ in eq.\ \eqref{SI:munabrho} and proceeding to the units of Active Model B where $\kappa(\phi)=1$ for $\phi=0$, the gradient term \eqref{SI:munabrho} transforms into eq.\ \eqref{SI:munab} with the linear function $\kappa(\phi)=1+2\lambda\phi$ and the negative density-independent parameter 
\begin{equation}
\lambda=\frac{\kappa'(\phi)}{2}=-\frac{1}{4}v(0)\sigma_{\mathrm{s}}\tau_{\mathrm{c}}\frac{\rho_{\mathrm{H}}-\rho_{\mathrm{L}}}{1-v(0)\sigma_{\mathrm{s}}\tau_{\mathrm{c}}\bar{\rho}} \;.
\label{SI:lambda0}%
\end{equation}%
This expression can also be written as
\begin{equation}
\lambda=\frac{\rho_{\mathrm{H}}-\rho_{\mathrm{L}}}{4\bar{\rho}}\frac{\dif\ln(v)}{\dif\ln(\rho)}\bigg\rvert_{\rho=\bar{\rho}} 
\label{SI:lambda}%
\end{equation}%
where $\dif\ln(v)/\dif\ln(\rho)$ and $(\rho_{\mathrm{H}}-\rho_{\mathrm{L}})/(4\bar{\rho})$ are each dimensionless numbers of order unity. Hence $\lambda$ is negative and of order unity in general.
These considerations as well as the direct simulations of ref.\ \cite{StenhammarTAMC2013} confirm the general results derived further above for the special case of ABPs and show how the parameter $\lambda$ is related to the underlying physics.
The same is true for run-and-tumble bacteria, whose dynamics are almost equivalent 
to ABPs \cite{CatesT2013}, except for the fact that $v(\rho)$ is now a more general function of $\rho$ than the linear form considered above for ABPs. (The relation between $\kappa_{\rho}(\rho)$ and $v(\rho)$ is however unchanged.)

Active Model B can now be written as the conservation equation 
\begin{equation}
\dot{\phi}+\nabla\!\cdot\!\mathbfup{J}=0
\label{SI:eq:AMB_ce}%
\end{equation}
with the current (suppressing the noise term $\mathbfup{\Lambda}$)
\begin{equation}
\mathbfup{J}=(1-3\phi^{2})\nabla\phi+\nabla^{3}\phi-\lambda\nabla(\nabla\phi)^{2} \;,
\label{SI:eq:AMB_J}%
\end{equation}
which together imply the Cahn-Hilliard-like equation
\begin{equation}
\dot{\phi}=\nabla^{2}\big(-\phi+\phi^{3}-\nabla^{2}\phi+\lambda(\nabla\phi)^{2}\big) \;.
\label{SI:eq:AMB}%
\end{equation}
Note that the sign of $\lambda$ can be absorbed, if desired, into that of $\phi$. 

We next show that all possible current contributions up to third order in $\nabla$ and second order in $\phi$ can be decomposed into an integrable part (i.e., can be written as $\nabla\delta\mathcal{F}/\delta\phi$ for some choice of $\mathcal{F}=\int\! f\,\dif^{d}r$) plus a linear combination of $\nabla(\nabla\phi)^{2}$ (which is our $\lambda$ piece) and the term $\phi\nabla^{3}\phi$ that can be removed by redefining the diffusivity.
To show this, it is sufficient to consider all terms that are of first or third order in $\nabla$ and of first or second order in $\phi$. (All other terms to the required order are not vector-valued and so cannot contribute to the current in eq.\ \eqref{SI:eq:AMB_J}.) 
Obviously, all terms that are of first order in $\nabla$ and of first or second order in $\phi$ are reducible to multiples of $\nabla\phi$ and $\phi\nabla\phi$. These terms are integrable with corresponding free-energy densities $\phi^{2}/2$ and $\phi^{3}/6$, respectively. 
There is only one vector-valued term that is of third order in $\nabla$ and of first order in $\phi$. 
This term $\nabla^{3}\phi$ is integrable, too, with free-energy density $-(\nabla\phi)^{2}/2$.
The remaining terms are of third order in $\nabla$ and of second order in $\phi$. All vector-valued terms of this order can be written as linear combinations of $\phi\nabla^{3}\phi$, $(\nabla\phi)\nabla^{2}\phi$, and $\nabla(\nabla\phi)^{2}$. 
Any corresponding free-energy density would be of second order in $\nabla$ and of third order in $\phi$. Such a free-energy density is scalar and can therefore only be a linear combination of $\phi(\nabla\phi)^{2}$ and $\phi^{2}\nabla^{2}\phi$; but these two terms are equivalent (modulo boundary terms) as can be shown by partial integration. The gradient of the functional derivative of any combination of them is therefore proportional to $\nabla(\nabla\phi)^{2}+2(\nabla\phi)\nabla^{2}\phi+2\phi\nabla^{3}\phi$. 
This means that one of the three terms $\phi\nabla^{3}\phi$, $(\nabla\phi)\nabla^{2}\phi$, and $\nabla(\nabla\phi)^{2}$ can be written as an integrable contribution, plus a linear combination of the two other terms which remain, in principle, nonintegrable and independent. Here we choose $\phi\nabla^{3}\phi$ and $\nabla(\nabla\phi)^{2}$ as the only two independent nonintegrable current contributions.

Importantly, however, there is already a term  $\nabla^{3}\phi$ in the current \eqref{SI:eq:AMB_J}. Moreover, multiplying the entire current by any function of $\phi$ is equivalent to introducing a density-dependent diffusivity. (Note that in Active Model B, as in the passive version, we have already ignored such a dependence and also its effect on the noise term $\mathbfup{\Lambda}$.) Therefore the $\phi\nabla^{3}\phi$ term in $\mathbfup{J}$ can be absorbed by combining a diffusivity linear in $\phi$ with a different choice of the bulk free-energy density $f_{0}(\phi)$, chosen to restore the form of the first term in eq.\ \eqref{SI:eq:AMB_J}. The final result of this process is to add to $\mathbfup{J}$ a new term proportional to $\phi\nabla(\nabla\phi)^{2}$ [equivalent to choosing a non-constant but linear $\lambda(\phi)$] which is of third order in both $\phi$ and $\nabla$, so we neglect it. This leaves, to this order, only one genuinely nonintegrable term, $\nabla(\nabla\phi)^{2}$, which we retain, justifying the choice $\mu^{\mathrm{A}}_{1} = \lambda(\nabla\phi)^{2}$, and thereby eq.\ \eqref{SI:eq:AMB}. 

The $\lambda$ term is thus confirmed as the unique leading-order TRS-breaking term for inclusion in Active Model B.

\section{Supplementary Note 2: Balancing pressures}
In this Supplementary Note we consider the existence and stability of a circular ($d=2$) or spherical ($d=3$) droplet as a static solution of Active Model B [see eq.\ \eqref{SI:eq:AMB}] for $\mu=0$. 
For convenience, we here assume that the droplet is located in the origin of coordinates at $\mathbfup{r}=\mathbfup{0}$ so that the stationary solution $\phi(r)$ has full rotational symmetry and depends only on the radial coordinate $r=\lvert\lvert\mathbfup{r}\rvert\rvert$. 
Equation \eqref{mu} then reduces to the one-dimensional ordinary differential equation (ODE) 
\begin{equation}
-\phi+\phi^{3}-\frac{d-1}{r}\phi'-\phi''+\lambda (\phi')^{2}=\mu \;.
\label{SI:eq:phiDroplet}%
\end{equation}
This ODE cannot be solved analytically, but for a large droplet with radius $R^{*}$ a perturbative solution can be derived.
Such solutions with $R^{*}\gg 1$ can be expected to exist for $\lvert\lambda\rvert\ll 1$, since we know from the static and stable solution of Passive Model B ($\lambda=0$), with one planar interface connecting two coexisting bulk phases, that $R^{*}\to\infty$ for $\lambda\to 0$.
In order to find a perturbative solution to leading order in large $R^{*}$, we replace $(d-1)/r$ in eq.\ \eqref{SI:eq:phiDroplet} by $(d-1)/R^{*}$. This is possible for droplets whose interface width is much smaller than their radius $R^{*}$,  so that $1/r$ is approximatively constant through the interface. Furthermore, we choose $\mu=0$, thereby restricting the search to `droplet solutions' of Active Model B (comprising a single large droplet in coexistence with an infinite bulk of the other phase) in which the Laplace pressure of the droplet and the influence of the $\lambda$ term balance each other. 
This results in 
\begin{equation}
-\phi+\phi^{3}-\xi\phi'-\phi''+\lambda (\phi')^{2}=0 
\label{SI:eq:phiXi}%
\end{equation}
with $\xi=(d-1)/R^{*}$. Notice the invariance of this equation with respect to a double sign switch ($\phi\to-\phi$, $\lambda\to-\lambda$), meaning that the sign of $\lambda$ only decides whether the droplet consists of the dense or the dilute phase.
It is therefore sufficient to consider only the case $\lambda\leqslant 0$ in the following. 
The parameters $\xi>0$ and $\lambda$ in eq.\ \eqref{SI:eq:phiXi} have now to be chosen in such a way that this equation has a `droplet solution' as defined above.

Using the Newton mapping ($z\to t$, $\phi\to x$) \cite{OxtobyH1982}, eq.\ \eqref{SI:eq:phiXi} can be written as the equation of motion
\begin{equation}
\ddot{x}=-U'(x)+\lambda \dot{x}^{2}-\xi\dot{x} 
\label{SI:eq:xDroplet}%
\end{equation}
of a Newtonian particle with mass $m=1$ in the symmetric external potential $U(x)=-f_{0}(x)$, where besides the time-reversal invariant force $\lambda\dot{x}^{2}$, a frictional force $-\xi\dot{x}$ is present. 
With the subsequent substitutions $\nu(x)=\dot{x}(t)$ and $w(x)=\nu^{2}(x)$, eq.\ \eqref{SI:eq:xDroplet} can be transformed into the nonlinear first-order ODE 
\begin{equation}
w'(x)=2\lambda w(x) -2\:\!\xi\sqrt{w(x)}-2\:\!U'(x) 
\label{SI:eq:wDroplet}%
\end{equation}
for the squared velocity $w(x)$ of the Newtonian particle. This equation cannot be solved analytically. 

Since the maxima of the potential $U(x)$ are at $x^{(1)}_{\mathrm{max}}=-1$ and $x^{(2)}_{\mathrm{max}}=1$, the solutions of eq.\ \eqref{SI:eq:wDroplet} can be classified as follows: 
i) solutions with positions $x_{1}$ and $x_{2}$ so that $-1\leqslant x_{1}\leqslant x\leqslant x_{2}<1$ or $-1<x_{1}\leqslant x\leqslant x_{2}\leqslant 1$ as well as $x_{1}<x_{2}$ and $w(x_{1})=w(x_{2})=0$ (these solutions describe a damped oscillation of the position $x(t)$ of the Newtonian particle), 
ii) solutions with $-1\leqslant x\leqslant 1$ and $w(-1)=w(1)=0$ that describe the non-oscillatory motion of $x(t)$ from $x^{(1)}_{\mathrm{max}}$ to $x^{(2)}_{\mathrm{max}}$ or from $x^{(2)}_{\mathrm{max}}$ to $x^{(1)}_{\mathrm{max}}$ -- depending on the sign of $\lambda$ (these are the required droplet solutions), 
and iii) unphysical solutions in all other cases. 
Evaluating eq.\ \eqref{SI:eq:wDroplet} numerically and considering only solutions of type ii) leads to the linear relation $\xi\approx 0.566\:\!\lvert\lambda\rvert$ between the parameters $\xi$ and $\lvert\lambda\rvert$. 

Below we confirm that the exact relation between $\xi$ and $\lambda$ is given by   
\begin{equation}
\xi=\frac{\sqrt{8}}{5}\lvert\lambda\rvert \;,
\label{SI:eq:xiDroplet}%
\end{equation}
where indeed $\sqrt{8}/5 \approx 0.566$. Thus solutions representing large droplets with $\mu=0$ do exist; in these solutions the Laplace pressure is compensated by the $\lambda$ term. The radius $R^{*}$ of such a droplet in $d$ spatial dimensions is then given as  
\begin{equation}
R^{*}=\frac{5}{\sqrt{8}}\frac{d-1}{\lvert\lambda\rvert} \;.
\label{SI:eq:RDroplet}%
\end{equation}
This result can be confirmed simply by balancing the Laplace pressure $\Delta P_{\mathrm{L}}$, caused by the interfacial tension at a curved interface, with the pressure jump $\Delta P_{\lambda}$ at the interface that originates from the $\lambda$ term. Indeed, the Laplace pressure of the droplet is given by 
\begin{equation}
\Delta P_{\mathrm{L}}=\gamma\frac{d-1}{R^{*}}
\label{SI:eq:PL}%
\end{equation}
with the standard interfacial tension $\gamma=\sqrt{8}/3$ of Model B \cite{KendonCPDB2001}.
On the other hand, the pressure difference $\Delta P_{\lambda}=P_{2}-P_{1}$ originating from the $\lambda$ term is the difference of the pressure $P_{2}$ inside the droplet and the pressure $P_{1}$ outside the droplet, where the pressure of a bulk phase is defined as $P=\phi\mu_{0}-f_{0}$. To calculate the active pressure we consider bulk phase coexistence across a planar interface (as we require this quantity only to zeroth order in $1/R^{*}$). Since in both coexisting bulk phases the nonequilibrium chemical potential has the same value $\mu_{\mathrm{b}}$, the pressure difference can be written as $\Delta P_{\lambda}=\mu_{\mathrm{b}}(\phi_{2}-\phi_{1})+f_{0}(\phi_{1})-f_{0}(\phi_{2})$, where $\phi_{1}$ and $\phi_{2}$ are the densities of the coexisting phases (chosen to match those outside and inside the droplet, respectively). 
For $|\lambda|$ small as assumed above $\phi_{1}\to-\phi_{2}$, $\lvert\phi_{1}\rvert,\lvert\phi_{2}\rvert\to 1$, and $\mu_{\mathrm{b}}\to 4\lambda/15$ [see eq.\ \eqref{SI:eq:mub}]. Hence the activity-induced pressure difference becomes $\Delta P_{\lambda}=8\lambda\phi_{2}/15$. 
The expression $\Delta P_{\lambda}=8\lambda\phi_{2}/15$ shows that for $\lambda<0$ only droplets of the dense phase surrounded by the dilute phase can exist, whereas for $\lambda>0$ only droplets of the dilute phase in the dense phase are possible. 
Since for a static droplet the signs of $\lambda$ and $\phi_{2}$ obviously have to be different so that $\Delta P_{\lambda}$ counteracts $\Delta P_{\mathrm{L}}$, the expression for the active pressure $\Delta P_{\lambda}$ can further be simplified to 
\begin{equation}
\Delta P_{\lambda}=-\frac{8}{15}\lvert\lambda\rvert \;.
\label{SI:eq:Plambda}%
\end{equation}
Balancing the pressures \eqref{SI:eq:PL} and \eqref{SI:eq:Plambda} by $\Delta P_{\mathrm{L}}+\Delta P_{\lambda}=0$ then results in eq.\ \eqref{SI:eq:RDroplet} for the radius $R^{*}$ of the droplet.


This result shows that there exists a static solution of Active Model B that describes a single droplet.
However, this solution is not stable. To show this, Active Model B \eqref{SI:eq:AMB} is first written in polar ($d=2$) or spherical ($d=3$) coordinates, to take advantage of the rotational symmetry. 
Denoting the static droplet solution as $\phi_{\mathrm{D}}(r)$, and adding a perturbation $\delta\phi(r,t)$, we find from eq.\ \eqref{SI:eq:AMB} a dynamical equation for the perturbation $\delta\phi(r,t)$.
We assume that the perturbation $\delta\phi(r,t)$ is very small and that its spatial variations are even smaller. 
The spatial derivatives of $\delta\phi(r,t)$ and all terms that are nonlinear in $\delta\phi(r,t)$ can then be neglected. 
This simplification results in the following dynamical equation for the perturbation: 
{\allowdisplaybreaks
\begin{gather}%
\begin{split}%
\!\!\!\!\!\!\!\!\!\!\!\!\dot{\delta\phi}(r,t)=s(r)\delta\phi(r,t) \;,\label{SI:eq:PerturbationA}%
\end{split}\\
\begin{split}%
s(r)=\frac{6}{r}\Big(r(\phi_{\mathrm{D}}')^{2}
+\phi_{\mathrm{D}}\big((d-1)\phi_{\mathrm{D}}'
+r\phi_{\mathrm{D}}''\big)\!\Big) \,.
\end{split}%
\label{SI:eq:PerturbationB}%
\end{gather}}%
From the Newtonian equation \eqref{SI:eq:xDroplet} it is clear that the droplet solution $\phi_{\mathrm{D}}(r)$ is a strictly monotonic function with a zero at $r=R^{*}$. We next consider eq.\ \eqref{SI:eq:PerturbationB} near the zero of $\phi_{\mathrm{D}}(r)$. At $r=R^{*}$ eq.\ \eqref{SI:eq:PerturbationB} reduces to $s(R^{*})=6(\phi_{\mathrm{D}}'(R^{*}))^{2}>0$ and the perturbation described by eq.\ \eqref{SI:eq:PerturbationA} grows linearly in time. The droplet solution of Active Model B is therefore unstable. 

Since eq.\ \eqref{SI:eq:PerturbationB} does not explicitly depend on $\lambda$, the above argument holds equally for Passive Model B. The droplet instability of Active Model B therefore is fundamentally no different from the classical instability of a fluid droplet with a certain stationary radius that is in \textit{unstable} equilibrium with its vapour.

\section{Supplementary Note 3: Ostwald ripening}
We now consider the effects of the $\lambda$ term on a fluid droplet with radius $R(t)$ of phase A that is surrounded by a continuous phase B in which further droplets of A are embedded far away. Apart from the new $\lambda$ term, our arguments are everywhere standard \cite{Cates2013}. For ease of comparison with these standard arguments, we treat the non-$\lambda$ physics as though it were entirely thermodynamic, and use the language of molecular solubility rather than $\phi^{4}$ theory. For this reason parameters such as the molecular volume now appear, but our conclusions, which concern qualitative stability issues rather than quantitative coarsening rates, are not affected by this change of viewpoint.
For simplicity we assume that $|\lambda|$ is small so that eq.\ \eqref{SI:eq:Plambda} can be used.

The distance of our droplet to other droplets of phase A is assumed to be large so that coalescence can be ruled out. 
Nonetheless, there is (for $\lambda=0$) a small equilibrium concentration $c^{\mathrm{eq}}_{\mathrm{A}}$ of species A in the nearly pure phase B that allows molecular diffusion of A between separated droplets. 
Treating the environment of a representative droplet as isotropic, the local concentration of species A in phase B can be described by a function $c_{\mathrm{A}}(r,t)$, where $r$ is the distance from the center of the A droplet that is under consideration. 
Just outside the droplet at $r=R^{+}(t)$, the concentration is given by 
\begin{equation}
c_{\mathrm{A}}(R^{+}\!,t)=c^{\mathrm{eq}}_{\mathrm{A}}\big(1+\beta V_{\mathrm{A}} \Delta P\big)
\label{SI:eq:cARp}%
\end{equation}
with $\beta$ the inverse thermal energy and $V_{\mathrm{A}}$ the molecular volume of A. Here $\Delta P(t)$ is the pressure excess interior to the droplet, and the dilute solution of A in B outside the droplet is treated as an ideal mixture.
The pressure difference $\Delta P(t)$ is given by $\Delta P(t) = \Delta P_{\mathrm{L}}(t)+\Delta P_{\lambda}$ with the Laplace pressure $\Delta P_{\mathrm{L}}(t)\propto \gamma/R(t)$ [see eq.\ \eqref{SI:eq:PL}] and the active pressure $\Delta P_{\lambda}\propto-\lvert\lambda\rvert$ [see eq.\ \eqref{SI:eq:Plambda}].
Hence, the reduced pressure $\Delta P(t)$, and therefore also the concentration $c_{\mathrm{A}}(R^{+}\!,t)$, are larger for a small droplet than for a big one; crucially this statement holds regardless of $\lambda$.
For this reason, there is a negative concentration gradient and thus a diffusive flux of A from small droplets to big droplets so that small droplets shrink while big droplets grow. This process is known as Ostwald ripening \cite{Ostwald1900}. 
Far away from the droplet, the concentration of A in B is given by 
\begin{equation}
c_{\mathrm{A}}(\infty,t)=c^{\mathrm{eq}}_{\mathrm{A}}\big(1+\varepsilon(t)\big) \;,
\label{SI:eq:cAinf}%
\end{equation}
where the supersaturation $\varepsilon(t)$ takes into account that the overall system of droplets (an emulsion) has not yet reached steady state. 

\begin{figure}
\centering
\includegraphics[width=\columnwidth]{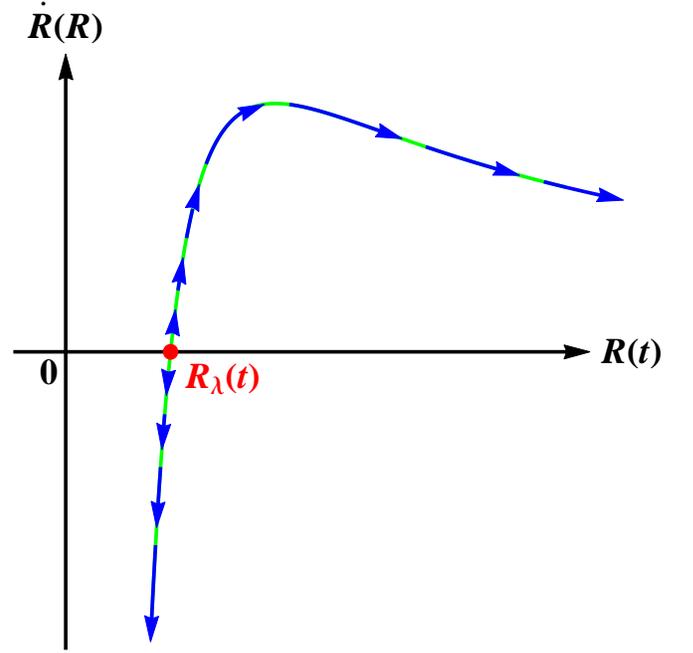}
\caption{\label{Fig.SI1}\AU{Sketch of the function $\boldsymbol{\dot{R}(R)}$.} There is an unstable fixed point at $R_{\lambda}(t)$. The arrows represent the positive or negative growth rate of a droplet with radius $R>R_{\lambda}$ or $R<R_{\lambda}$, respectively.\vspace*{9mm}}
\end{figure}

The dynamics of the concentration field $c_{\mathrm{A}}(r,t)$ is given by the diffusion equation $\dot{c}_{\mathrm{A}}+\partial_{r} J_{\mathrm{A}}=0$ with the current $J_{\mathrm{A}}(r,t)=-D_{\mathrm{A}}\partial_{r} c_{\mathrm{A}}(r,t)$, where $D_{\mathrm{A}}$ denotes the diffusion coefficient of A in B.
The solution of this diffusion equation with the boundary conditions \eqref{SI:eq:cARp} and \eqref{SI:eq:cAinf} is given by
\begin{equation}
c_{\mathrm{A}}(r,t)=c_{\mathrm{A}}(\infty,t)+\frac{R}{r}\big(c_{\mathrm{A}}(R^{+}\!,t)-c_{\mathrm{A}}(\infty,t)\big) \;.
\label{SI:eq:cA}%
\end{equation}
The temporal rate of change $\dot{R}(t)$ of the droplet radius is proportional to the current at its surface, $J_{\mathrm{A}}(R,t)=D_{\mathrm{A}}c^{\mathrm{eq}}_{\mathrm{A}}(\beta V_{\mathrm{A}} \Delta P(t)-\varepsilon(t))/R$, so that 
\begin{equation}
\dot{R}\propto \frac{1}{R}\bigg(\varepsilon_{\lambda}-\beta V_{\mathrm{A}}\gamma\frac{d-1}{R}\bigg) \,,
\label{SI:eq:R}%
\end{equation}
where we have introduced the shifted supersaturation 
\begin{equation}
\varepsilon_{\lambda}(t)=\varepsilon(t)+\beta V_{\mathrm{A}} \gamma\frac{d-1}{R^{*}(\lambda)} \;.
\label{SI:eq:epsLambda}%
\end{equation}
The function $\dot{R}(R)$ is zero at
\begin{equation}
R_{\lambda}(t)=\beta V_{\mathrm{A}}\gamma\frac{d-1}{\varepsilon_{\lambda}(t)} \;,
\label{SI:eq:Rlambda}%
\end{equation}
negative for $R<R_{\lambda}(t)$, and positive for $R>R_{\lambda}(t)$.
This means that the radius $R(t)$ has an unstable fixed point at $R_{\lambda}(t)$ (see Fig.\ \ref{Fig.SI1} for a sketch). 
A droplet with this radius is metastable, whereas smaller droplets shrink and bigger droplets grow. 
The metastable radius $R_{\lambda}(t)$ increases with time. 


Crucially, however, this description of the Ostwald process differs from the standard one (which applies at $\lambda = 0$) \textit{only} through the shifted supersaturation defined in eq.\ \eqref{SI:eq:epsLambda}. 
In the standard process ($\lambda=0$), the droplet radius corresponding to the unstable fixed point, $R_{0}(t)$, tends to infinity like $t^{1/3}$ at late times, while $\varepsilon_{0}(t)=\varepsilon(t)\to 0$. The same must therefore apply here ($\lambda\neq 0$), except that now $\varepsilon_{\lambda}(t)\to 0$, which means that the final supersaturation $\varepsilon(\infty)$ is not zero.

Indeed, by assuming that the typical droplet size $R_{\mathrm{m}}(t)$ is comparable, but not equal, to $R_{\lambda}(t)$, one can directly derive the temporal scaling of $R_{\mathrm{m}}(t)$, as
\begin{equation}
\dot{R}_{\mathrm{m}}\propto \beta V_{\mathrm{A}}\gamma\frac{d-1}{R^{2}_{\mathrm{m}}}
\label{SI:eq:RmODE}%
\end{equation}
from eq.\ \eqref{SI:eq:R}. Here $\lambda$ does not enter, and the result is the standard Ostwald scaling law, 
$R_{\mathrm{m}}(t)\propto t^{1/3}$. 
(With eq.\ \eqref{SI:eq:Rlambda} this implies the scaling law $\varepsilon_{\lambda}(t)\propto t^{-1/3}$.) 
Since this scaling argument gives the correct asymptotic coarsening law for Passive Model B (regardless of whether the morphology is that of discontinuous droplets as assumed, or bicontinuous), it compellingly suggests the same asymptotic scaling applies in Active Model B. If so, deviations in the apparent exponents from $1/3$, reported in Fig.\ \ref{Fig.2}, are transients rather than genuinely new asymptotic behavior.

\begin{figure}
\centering
\includegraphics[width=\columnwidth]{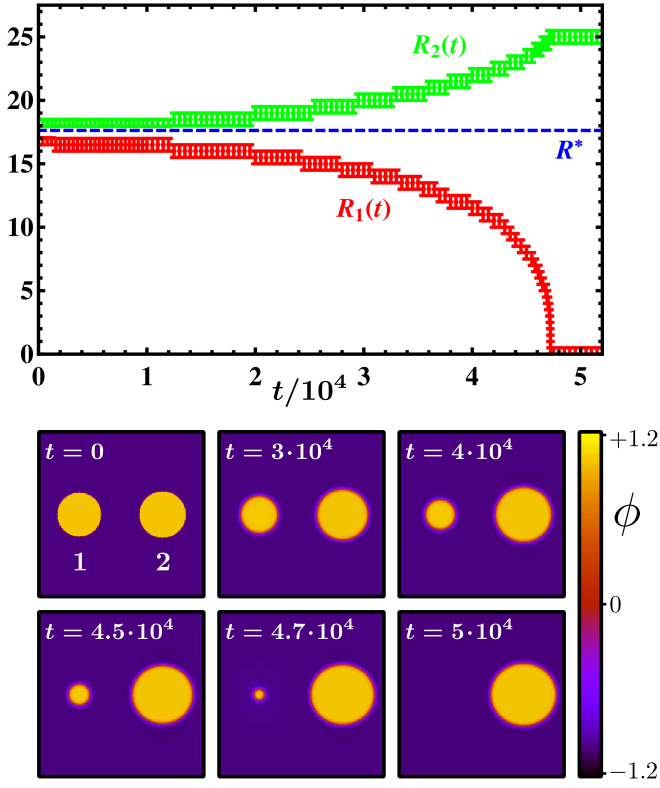}
\caption{\label{Fig.SI2}\AU{Ostwald ripening of two big droplets.} This plot shows the time evolution of two droplets with the slightly different initial radii $R_{1}(0)=17$ and $R_{2}(0)=18$ for $\lambda=-0.1$. The metastable radius at $\mu=0$ is $R^{*}(\lambda)\approx 17.7$ and the error bars denote the spatial discretization error.}
\end{figure}

By confirming that the nonintegrable $\lambda$ term cannot prevent Ostwald ripening, these arguments also show that all stable stationary states involve bulk phase separation rather than micro-phase separation into droplet phases. This is despite the fact that $R_\lambda = R^{*}(\lambda)$ when $\varepsilon=0$, which might appear to suggest a final state of stable droplets of size $R^{*}$ with $\varepsilon(t)\to 0$ at late times. (As shown above, it is $\varepsilon_{\lambda}(t)$ and not $\varepsilon(t)$ that vanishes at late times.)
As a numerical example, the Ostwald ripening of two large droplets with initial radii close to $R^{*}(\lambda)$ is shown in Fig.\ \ref{Fig.SI2}.
In this simulation, the smaller droplet shrinks until it vanishes, while the slightly bigger droplet grows.

\end{document}